\documentclass[11pt,twoside]{article}
\usepackage{jltp}
\usepackage{epsfig}
\usepackage{amssymb}

\def\bigPi{\mathord{\mbox{\large $\Pi$}}}

\title{Theory for underdoped high-T$_{\rm c}$ superconductors: effects
  of phase fluctuations}

\author{F. Sch\"afer, C. Timm, D. Manske, and
  K.H. Bennemann\address{Institut f\"ur Theoretische Physik, Freie
  Universit\"at Berlin, Arnimallee 14 \\ D-14195 Berlin, Germany}}

\runninghead{F. Sch\"afer {\it et al.}}{Theory for underdoped
  high-T$_{\rm c}$ superconductors}
       
\begin{document}

\begin{abstract}
In underdoped cuprates, $T_{\rm c}$ is thought to be determined by
Cooper pair phase fluctuations because of the small superfluid
density $n_s$. Experimentally, $T_{\rm c}$ is found to scale with
$n_s$. The fluctuation-exchange approximation (FLEX) in its standard
form fails to predict this behavior of $T_{\rm c}$ since it does not
include phase fluctuations. We therefore extend the FLEX
to include them selfconsistently. We present results for $T_c[n_s,x]$,
where $x$ is the doping.   
 
PACS numbers: 74.72-h, 74.20.Mn, 74.25.Jb\hspace*{-4cm}
\end{abstract}

\maketitle
\vspace{0.3in}
One of the striking properties of underdoped cuprates is the 
so-called strong pseudogap above $T_c$ \cite{Timusk}. 
It appears below a characteristic temperature $T^*_c$
in the underdoped regime and $T^*_c$ approaches the superconducting
transition temperature $T_c$ near optimal doping, where T$_c$ is maximal. 
The pseudogap evolves smoothly into the superconducting gap and has
the same {\it d}-wave symmetry. 
A further characteristic property of underdoped cuprates is the low
superfluid density $n_s$. 
Furthermore, $T_c \propto n_s(T=0)/m^*$ is 
observed experimentally \cite{Uemura} and there are indications of
finite phase stiffness being proportional to $n_s$ on short 
length and time scales in the pseudogap regime above $T_c$ \cite{Corson}.  
One interpretation is that at $T_c^*$ Cooper pair formation occurs and
that at $T_c$ these Cooper pairs become phase coherent.
In accordance with this physical picture we extend FLEX to include
phase fluctuations.   

\section*{Theory}
As a microscopic model for  high-T$_c$ superconductors we use the
single-band Hubbard model in 2D, with the nearest-neighbor
hopping element $t$ and the local Coulomb repulsion $U$. The
fluctuation exchange approximation \cite{Bickers} (FLEX) is used to
treat the superconducting state in the standard Nambu-Eliashberg
formalism with the spin fluctuations of the FLEX as the pairing
interaction \cite{Monthoux,Dahm}.   
From the self-consistent solution of the Dyson
equation one can determine $T^*_c$ as the highest temperature for
which the off-diagonal self energy is non-vanishing.
Usually this is the criterion for the critical temperature
$T_c$, but this is only true if one can neglect the role of phase
fluctuations. However, when the superfluid density is small, phase
fluctuations will destroy phase coherence, while the charge carriers
still form local Cooper pairs, which is signaled by a non zero
amplitude of the superconducting order parameter. 

\begin{figure}
\centerline{\psfig{file=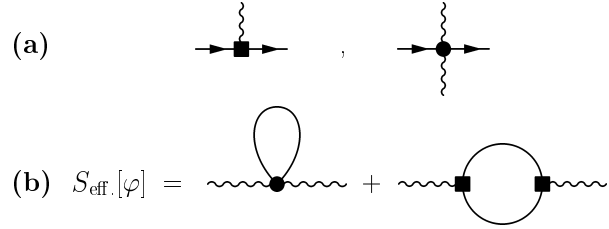,width=9cm,bb=117 568 533 736}}
\caption{\it In (a) the elementary vertices {\small $\blacksquare$}
  and {\LARGE $\bullet$} of our theory are 
  shown. The solid line represent an in or outgoing electron and the
  wiggly the phase field.  In (b) we show the diagrams contributing in
  lowest order to the effective action ${\cal S}_{eff.}[\varphi]$ for
  the phases, here the solid line is the Nambu Green function.}  
\label{fig:vertices} 
\end{figure}

To extend the FLEX to include phase fluctuations
self-consistently, one has to know their self-energy contribution.
In order to derive it we start with a phenomenological model 
of tight-binding electrons and an effective nearest-neighbor
pairing interaction for electrons with opposite spins, leading to a
superconducting {\it d}-wave order parameter in the mean-field
approximation. 
We begin by writing down the action ${\cal S}$ for the tight-binding
electrons where the interaction term is decoupled by a
Hubbard-Stratonovich transformation, introducing a decoupling
field $\Delta_{i j}$, which turns out to be the superconducting order
parameter:
\begin{eqnarray}
  {\cal S}[\Phi^*,\Phi;\Delta^*,\Delta]&=&
  \int_0^{\beta}d\tau\; \Bigg\{ \sum_{i \sigma} \Phi_{i
    \sigma}^* (\partial_{\tau} - \mu) \Phi_{i \sigma} + t \sum_{<i j>
    \sigma } \Phi^*_{i \sigma} \Phi_{j \sigma} \nonumber \\
  &&+ \sum_{<i j>} \bigg[\Delta_{i j}\;\Phi^*_{i \uparrow} \;
  \Phi^*_{j \downarrow}+\Delta^*_{i j} \; \Phi_{j \downarrow} \;
  \Phi_{i \uparrow} + \frac{|\Delta_{i j}|^2}{|V|^{\phantom{2}}} \bigg]
  \Bigg\} \quad\mbox{,} 
  \nonumber\label{Wirkung1}
\end{eqnarray}
where $\Phi_{i \sigma}$ represents the electron field at lattice
site $i$ with spin $\sigma$. $V$ is the pairing interaction and $t$ is 
the hopping matrix element between nearest-neighbor sites. 
The doping is controlled by the chemical potential $\mu$. 
We now assume the local superconducting order parameter $\Delta_{i
  j}(\tau)$ to have a time and translational invariant amplitude
$\Delta^0_{i j}$ with {\it d}-wave symmetry but a fluctuating phase
suppressing superconducting order in the pseudogap regime below
$T_c^*$.  
We perform the following transformation to decouple the phase
and amplitude degrees of freedom of the order parameter \cite{Kwon}:
\begin{eqnarray}
  \psi_{i \sigma} = \Phi_{i \sigma} \; {\rm
    e}^{-\frac{i}{2}\varphi_i(\tau)} \quad\mbox{,}\qquad
  \Delta_{i j}\ = \Delta^0_{i j} \; {\rm
    e}^{-\frac{i}{2}[\varphi_i(\tau)+\varphi_j(\tau)]}
  \quad\mbox{.}\nonumber
\end{eqnarray}
This leads to an expression for the action where the phase is directly
coupled to the fermions by the elementary vertices shown in figure
\ref{fig:vertices}.
After integrating out the fermionic degree of freedom, a loop
expansion up to second order in the phase fields $\varphi_i(\tau)$
leads to the effective action for the phases, taking the
form of a time-dependent XY model on a cubic lattice. From the
effective action we determine the propagator of the phase
fluctuations:    
\begin{eqnarray}
  \bigPi^{\varphi \varphi}(q,i\nu_n) = \frac{8}{-\kappa
  (q,i\nu_n) \; (i\nu_n)^2 \;+\; \frac{n_s(q,i\nu_n)}{m^*}
  \; 2(2- \cos q_x - \cos q_y)}  \quad\mbox{.}
\end{eqnarray}
where $\kappa(q,i\nu_n)$ is the density-density correlation function
and the dynamical phase stiffness is given by ${n_s}(q,i\nu_n)/{m^*} =
{\rm T} + \Pi^{jj}(q,\omega)/{e^2}$. 
Here ${\rm T}$ is the
expectation value of the inverse band-mass operator and $\bigPi^{jj}$
is the current-current correlation function calculated in the
mean-field approximation.  
The superfluid density is the density of {\it phase coherent}
Cooper pairs only, which are responsible for the Meissner effect.
If no phase fluctuations are present it will take
the  value of the local Cooper-pair density $n_{cp}$. Since in the
London theory $n_s$ is related to the London penetration depth by
$\lambda_L = \sqrt{m^* c^2/(4\pi e^2 n_s)}$ we can determine $T_c$ as
the temperature where the Meissner effect and therefore the superfluid
density $n_s(q\!\!=\!\!0,i\nu_n\!\!=\!\!0)$ refering to phase
coherent Cooper pairs vanishes. This can be seen
as the loss of rigidity against long-range phase fluctuations and is
consistent with the transition from quasi long-range to short-range
order in 2D \cite{Rice}. 
For $T>T_c$ the dynamical phase stiffness will be non-zero for certain
$q>q_0$ and $\nu>\nu_0$ and will result in superconducting order on
finite length and time scales and will finally vanish for all momenta
and frequencies at $T_c^*$, where $n_{cp}$ goes to zero \cite{Corson}.  
\begin{figure}
\centerline{\psfig{file=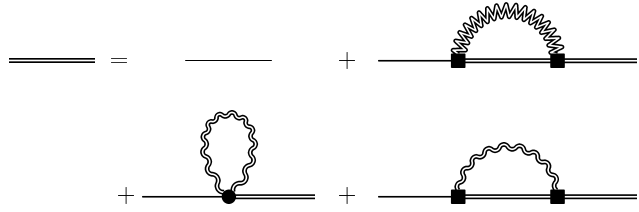,width=9cm,bb=70 545 515 705}}
\caption{\it Dyson equation for the extended FLEX approximation.
The straight double lines are renormalized Nambu Green
function, whereas the single lines are bare ones. The zig-zag line
in the second graph is the effective interaction of the FLEX, and the
wiggly lines denote the propagator $\Pi^{\varphi \varphi}$ of the
phase fluctuations.}  
\label{fig:dyson} 
\end{figure}
From the generating functional for the fermions we determine
the self-energy corrections due to phase fluctuations in lowest order.
Combining this self-energy contribution with the contribution of standard FLEX
and taking the renormalized propagator of the phase fluctuations,
which means that one has to calculate $n_s(q,i\nu_n)$ and
$\kappa(q,i\nu_n)$ with  {\it renormalized} Green functions, gives a set of
self-consistency equation for the renormalized Green functions.
In this way we get an extension of the usual FLEX equations with the
Dyson equation shown in figure \ref{fig:dyson}. 

\section*{Results}

\begin{figure}
\centerline{\psfig{file=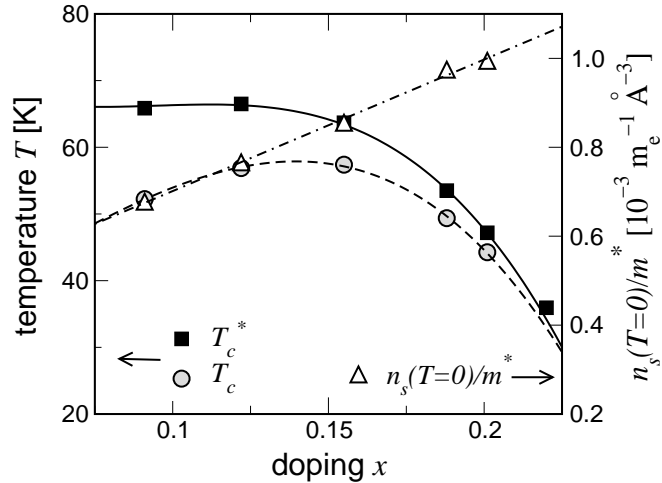,width=6.6cm,clip=
    ,angle=-90,bb=75 0 530 620}}
\caption{\it Doping dependence of the temperatures $T_c^*$ where
  Cooper pairs are formed, of $T_c$ where Cooper pairs becomes phase
  coherent, and of the phase stiffness $n_s(T=0)/m^*$. $T_c^*$ and
  $n_s(T)/m^*$ are calculated using FLEX and $T_c$ is calculated using
  the Kosterlitz-Thouless criterion for the vanishing of the phase
  coherence.} 
\label{fig:phasediagram} 
\end{figure}

In figure \ref{fig:phasediagram} we show results for the critical
temperature $T_c^*$ at which Cooper pairs are formed and for $T_c$
where the phase of the Cooper pairs become coherent.
$T_c^*$ is determined within standard FLEX as the highest temperature
for which the off-diagonal self energy is non-zero.
$T_c$ is calculated using the Kosterlitz-Thouless
criterion\cite{Kosterlitz} $\pi/2\: \hbar^2 d \: n_s(T_c)/4m^* = k_B
T_c$ for the temperature where phase coherence vanishes. Here
$n_s(T)/m^*$ is calculated within FLEX approximation and $d$ is half
of the $c$-axis lattice constant.  
Note the Kosterlitz-Thouless criterion neglects the coupling between
amplitude and phase fluctuations, which is expected to be important
for the overdoped regime. 
Also we show FLEX results for $n_s/m^*$ extrapolated to $T=0$,
indicating $T_c\propto n_s(T=0)/m^*$ in the underdoped regime. We find 
$T_c=1.691 \hbar^2 d \: n_s/4m^*\:k_B$ \cite{3DXY}.
In accordance with experimental findings we expect that our results for
$n_s(T=0)/m^*$ and $T_c$ should decrease more rapidly for doping
$x \rightarrow 0$ due to the vicinity of the antiferromagnetic transition
which is not properly described in the FLEX. 
It is interesting to note, that $T_c \rightarrow T_c^*$ in the
overdoped regime as expected for increasing Cooper pair density,
although using the Kosterlitz-Thouless criterion neglects coupling of
amplitude and phase fluctuations. We obtain $T_c \approx T_c^*$, since
$n_s/m^*$ increases rapidly below $T_c^*$ as compared to the
underdoped regime. 
The result for $n_s(T=0)/m^*$ in the overdoped regime suggests that
$T_c\propto n_s(T=0)/m^*$ is no longer valid there.
The scaled curve $T_c \propto n_s(T=0)/m^*$ crosses $T_c^*$ at optimal
doping.  
Our result for $T_c$ and $T_c^*$ already yield the observed behavior.
We expect that a selfconsistent inclusion of
phase fluctuations in FLEX will give better agreement with experiment, 
in particular a more rapid decrease of $T_c$ for $x \rightarrow 0$. 
The extended FLEX theory shown in figure \ref{fig:dyson} should give
for the underdoped superconductors $T_c \propto n_s$,
$n_s(q\!\!=\!\!0,i\nu_n\!\!=\!\!0)=0$ at $T_c$ to obtain no Meissner
effect above $T_c$, and the pseudogaps in the 
single-particle  spectral density. Note, the coupling of the phase
to its conjugate variable, the 
charge density, becomes important as the doping goes to zero.
This may cause $T_c \rightarrow 0$ for $x \rightarrow 0$.

{\bf Acknowledgments:} One of us (D.M.) would like to thank T. Dahm
for useful discussions and numerical help.


\begin{thebibliography}{9}
\bibitem{Timusk} For a recent review see, {\it e.g.} T. Timusk and B. Statt
  , {\it Rep. Prog. Phys.} {\bf 62}, 61 (1999). 
\bibitem{Uemura} Y.~J. Uemura {\it et al.}, {\it Phys. Rev. Lett.}
  {\bf 62}, 2317 (1989). 
\bibitem{Corson} J. Corson {\it et al.}, {\it Nature} {\bf 398}, 221
  (1999).   
\bibitem{Emery} V.~J. Emery and S.~A. Kivelson, {\it Nature} {\bf
  374}, 434 (1995).
\bibitem{Bickers} N.~E. Bickers and D.~J. Scalapino, {\it
  Ann. Phys. (N.Y.)} {\bf 193}, 206 (1989).
\bibitem{Monthoux} P. Monthoux and D.~J. Scalapino, {\it
  Phys. Rev. Lett.} {\bf 72}, 1874 (1994).
\bibitem{Dahm} T. Dahm and L. Tewordt, {\it
  Phys. Rev. Lett.} {\bf 74}, 793 (1995). 
\bibitem{Kwon} H.-J. Kwon and A.~T. Dorsey, {\it Phys. Rev. B} {\bf
  59}, 6438 (1999). 
\bibitem{Rice} T.~M. Rice, {\it Phys. Rev.} {\bf 140}, A1889 (1965).
\bibitem{Kosterlitz} J.~M. Kosterlitz and D.~J. Thouless, {\it
  J. Phys. C} {\bf 6}, 1181 (1993).
\bibitem{3DXY} 3D isotropic XY model gives instead of $1.691$ the
  factor $2.202$.
\end{thebibliography}
\end{document}